\documentstyle[11pt,aaspp]{article}
\input epsf.sty
\begin{document}
\title{The Optical Gravitational Lensing Experiment.\\
Variable Stars in Globular Clusters -II. Fields 5139D-F in 
Omega Centauri
\footnote{
Based on observations collected at the Las Campanas Observatory of the
Carnegie Institution of Washington.
}}
\author{J.~Kaluzny, M.~Kubiak, M.~Szyma{\'n}ski, A.~Udalski}
\affil{Warsaw University Observatory, Al. Ujazdowskie ~4, 
00-478 Warsaw, Poland\\
e-mail: (jka,mk,msz,udalski)@sirius.astrouw.edu.pl}
\and
\author{W.~Krzemi{\'n}ski}
\affil{Carnegie Observatories, Las Campanas Observatory, Casilla 601,
La~Serena, Chile\\ 
e-mail: wojtek@roses.ctio.noao.edu}
\and
\author{Mario Mateo}
\affil{Department of Astronomy, University of Michigan, 821 Dennison
Bldg., Ann Arbor, MI 48109-1090, USA\\
e-mail: mateo@astro.lsa.umich.edu}
\and
\author{K. Stanek}
\affil{Princeton University Observatory, Peyton Hall, Princeton NJ 08544
USA\\
e-mail: stanek@astro.princeton.edu}
\begin{abstract}
Three fields located close to the center of the globular cluster 
$\omega$~Cen were surveyed in a search for variable stars.
We present V-band light curves for 32 periodic variables. This sample
includes 10 SX~Phe
stars, 19 eclipsing binaries, and three likely spotted variables 
(FK~Com or RS~CVn type stars). Only 4 of these 
variables were previously known (including two objects reported 
in Paper I). All SX~Phe stars and 8 eclipsing
binaries from our sample belong to blue stragglers.
Five binaries occupy positions on the upper main-sequence
of the cluster and another four are located at the base of the 
subgiant branch. 
Of particular interest is detection of a 
detached binary system with a period $P=4.64$ day. 
Further study of this star can provide direct information about
properties of turnoff stars in $\omega $~Cen. 
A detached or semi-detached system located
slightly above the horizontal branch of $\omega $~Cen was also 
discovered. 
All SX~Phe stars and most of eclipsing binaries discovered in our
survey are likely cluster members. 
We present $V$ vs. $V-I$ color-magnitude diagrams for the monitored
fields. A populous group of likely hot subdwarfs as well as
numerous candidates for blue stragglers can be noted in these 
diagrams. Our data indicate that $\omega $~Cen possesses the largest number of
blue stragglers among all galactic globulars. 
\end{abstract}
\section{Introduction}
The Optical Gravitational Lensing Experiment (OGLE) is a long term
project with the main goal of searching for dark matter in our Galaxy by
identifying microlensing events toward the galactic bulge (Udalski et
al. 1992, 1994). At times  the Bulge is unobservable we conduct other
long-term photometric programs.
A complete list of side-projects attempted by the OGLE team
can be found in Paczy{\'n}ski et al. (1995).
In particular, during last three observing seasons we monitored globular
clusters NGC~104 (=47~Tuc) and  NGC~5139 (=$\omega$~Cen) in a search for
variable stars of various types. Of primary interest was detection of 
detached eclipsing binaries. In Paper I (Kaluzny et al. 1996) we
presented results for three fields in $\omega$~Cen. Here we report on
variables discovered in another three fields of that cluster.

\section{Observations and data reduction}

The OGLE project is conducted using the 1-m Swope telescope at Las
Campanas Observatory which is operated by Carnegie Institution of
Washington. A single $2048\times 2048$ pixels Loral CCD chip, giving the
scale of 0.435 arcsec/pixel is used as the detector. The initial
processing of the raw frames is done automatically in near-real time.
Details of the standard OGLE processing techniques are described by
Udalski et al. (1992).

This paper is based mostly on the data obtained during 1996
observing season. The cluster was monitored for a period spanning
about 3 months. Detailed logs of observations can be found in Udalski
et al. (1995, 1996). We monitored three fields. Fields 5139D 
and 5139E were centered about 13 arcmin West and 
about 13 arcmin East of the cluster center, respectively. 
Field 5139F was centered about 12 arcmin South in respect to field 
5139E. The equatorial coordinates of centers of fields 5139D-F are given in
Table 1. Fields 5139E and 5139F were observed only in 1995. Field
5139D was additionally monitored for about 5 hours on the night of 21
June, 1995. 
%
%Table 1
\setcounter{table}{0}
\begin{table}
%\scriptsize
\caption[]{Equatorial coordinates for centers of 5139D-F fields.}
\begin{flushleft}
\begin{tabular}{lll}
\hline\noalign{\smallskip}
Field & $\alpha _{2000}$ & $\delta _{2000}$ \\
\hline\noalign{\smallskip}
5139D & 13:25:21.9 & -47:28:50 \\
5139E & 13:28:02.0 & -47:28:52 \\
5139F & 13:28:02.3 & -47:40:57 \\
\hline\noalign{\smallskip}
\hline\noalign{\smallskip}
\end{tabular}
\end{flushleft}
\end{table}
%%%%%%%%%
Most of the monitoring was performed through the Johnson V filter. 
Some exposures in the Kron-Cousins I band were also obtained. 
Table~2 gives total numbers of frames reduced for each of surveyed fields.
Most of observations in the V-band were collected with an exposure time
set to 420 seconds. For the I-band exposures lasted 300 seconds.  
For majority of the analyzed frames seeing was better than 1.6 arcsec. 
%Table 2
\setcounter{table}{1}
\begin{table}
%\scriptsize
\caption[]{Number of frames reduced for the $\omega $ Cen fields}
\begin{flushleft}
\begin{tabular}{lll}
\hline\noalign{\smallskip}
Field & Filter V & Filter I \\
\hline\noalign{\smallskip}
5139D & 233 & 36 \\
5139E & 282 & 44 \\
5139F & 139 & 10  \\
\hline\noalign{\smallskip}
\hline\noalign{\smallskip}
\end{tabular}
\end{flushleft}
\end{table}
The reduction techniques as well as algorithms used for selecting
potential variables are described in Paper I. The profile photometry 
was extracted with the help of DoPHOT (Schechter et al. 1993). 
The total number of stars contained in data bases with V photometry
was  37771, 49426 and 22115 for fields, 5139D, 5139E and 5139F, 
respectively. Most of these 
stars were faint objects with relatively large errors of photometry.
Table 3 gives condensed information about the numbers of stars analyzed for
variability and about the quality of derived photometry.
\section{Variable stars}
In this paper we present results for all variables identified in
observed fields with the exception of RR Lyr stars. 
RR Lyr stars in $\omega $~Cen shall be subject of a separate contribution. 
We identified a total of 32 periodic variables of which 28 are new discoveries. 
The rectangular and equatorial coordinates of these stars 
are listed in Table 4. 
The rectangular coordinates correspond to
positions of variables on the V-band "template" images which were 
submitted to the editors of A\&A (see Appendix A). These images 
allow easy identification of all objects listed in Table 4. 
The name of field in which a given variable can be 
identified is given in the 6th column. 
All frames collected by the OGLE team were deposited at the NASA NSS
Data Center \footnote{The OGLE data (FITS images) are accessible for
astronomical community
from the NASA NSS Data Center. Send e-mail to:
archives@nssdc.gfc.nasa.gov with the subject line: REQUEST OGLE ALL and 
put requested  frame numbers (in the form MR00NNNN where NNNN stands for
frame number according to OGLE notation), one per line, in the body
of the message. Requested frames will be available using 
an "anonymous ftp" service from nssdc.gfc.nasa.gov host in 
location shown in the return e-mail message from 
archives@nssdc.gsfc.nasa.gov
}. 
Frames mr10525, mr10524 and mr11042 were used as templates for 
fields 5139D, 5139E and 5139F, respectively.   
The transformation from rectangular to equatorial 
coordinates was derived based on positions of stars which could be 
matched with objects from the Guide Star Catalogue (Lasker et al. 1988).
We identified 77, 106 and  82 GSC stars 
for fields 5139D, 5139E  and 5139F, respectively. The adopted frame 
solutions reproduce equatorial coordinates of GSC stars with residuals 
rarely exceeding 0.5 arcsec. 

Our sample of variables consist of 10 SX~Phe stars and 19
certain or probable eclipsing binaries. 
Objects OGLEGC17 and OGLEGC20 were already reported in Paper I.
These two stars are located in the overlapping areas of fields 5139A/E
and 5139C/F. They were "re-discovered" in the photometry analyzed here.
Variables OGLEGC42=E39  and OGLEGC50=NJL220 were known before 
our survey (Jorgensen \& Hansen 1984). 

Table 5 lists basic characteristics of the light curves of 10 SX~Phe
stars identified in fields 5139D-F. The mean V magnitudes were calculated
by numerically integrating the phased light curves after converting them
into intensity scale. Photometric data for the remaining variables 
are given in Table 6.
To determine the most probable periods we used an {\it aov} statistic
(Schwarzenberg-Czerny 1989, 1991). This statistic allows -- in 
particular -- reliable determination of periods for variables with 
non-sinusoidal light curves (eg. eclipsing variables of EA-type).
Phased light curves of SX~Phe stars are shown in Fig. 1 while 
Figs. 2 and 3 present light curves for the remaining variables. 

Figure 4 shows location of variables with known color on the cluster 
color-magnitude diagram (CMD). 
For the SX~Phe stars marked positions correspond to the intensity-averaged
magnitudes. For the remaining variables we marked positions corresponding 
to magnitude at maximum light. The main sequence and red giant branch 
are marked with
double lines in Fig. 4. The large width of these sequences is due to the
well known spread of metallicity exhibited by stars in $\omega $~Cen
(Woolley 1966).

All SX~Phe stars as well as some eclipsing binaries are located on 
the cluster CMD among candidate blue stragglers. 
Four eclipsing binaries occupy positions at the base of the subgiant 
branch. Among them there are two systems with EA-type light 
curves: OGLEGC17 and OGLEGC51. The first system 
has been already reported in Paper I. The second one is a new discovery.
The orbital period of OGLEGC51, $P=4.64$ day, connected with the shape
of its light curve indicate that this binary is a detached system.
The phase coverage at the primary minimum is poor and observed
depth of the primary eclipse should be treated as a lower limit only.
The variable was caught twice in the primary eclipse and twice in the 
secondary eclipse. This allowed to determine the period of OGLEGC51
with confidence.

Five variables occupy positions on the upper part of the main-sequence of 
the cluster. OGLEGC53 exhibit EB-type light curve and has period close to 
0.50 day. It may be either a semidetached or nearly contact binary.
OGLEGC49, OGLEGC54 and OGLEGC56 show light curves and periods typical 
for W UMa-type contact binaries. OGLEGC43 also shows EW-type light curve  
but at the same time its period, $P=1.10$ day, is unexpectedly long for 
a contact binary of late spectral type. The light curve of this star 
shows two distinct minima. Hence, we can rule out a possibility
that OGLEGC43 is a pulsating variable with $P=0.55$ day
located in the galactic halo behind the cluster.

Another interesting binary identified in our survey is OGLEGC52.
This star is located slightly above the horizontal branch of
the cluster near the blue edge of the instability strip. 
Its light curve suggest a semi-detached configuration with  
a red component filling its Roche lobe. The secondary 
eclipse is  shallow but the depth of the primary eclipse 
indicates comparable sizes of components of the system. 
Most probably OGLEGC52 is composed of a horizontal branch star
and a red sub-giant. The out-of-eclipse part of the light curve does 
not show presence of any secondary periodicity. Hence, the blue
component of OGLEGC52 is not a RR~Lyr star. Some scatter visible in the
light curve before primary eclipse is due to problems with photometring
over-exposed images of the star. 

We consider some of our period determinations, especially those for
OGLEGC61 and OGLEGC65, as preliminary. For OGLEGC61 we obtained
reasonable, low-amplitude light curves for two values of the orbital 
period: 0.631 and 0.479 day. For OGLEGC65 the possible periods are
0.512 and 0.408 day.

The periods of all identified SX~Phe stars are rather firmly
established. For variables of this type we calculated the power spectra
using the program based on a CLEAN algorithm (Roberts et al. 1987). 
For all stars the derived power spectra show a very prominent peak at
positions corresponding to the periods listed in Table 5.

The variable OGLEGC41 is located about 0.3 mag to the red of the
subgiant branch of the cluster. We adopted for this star a period 
$P=1.62$ day but it cannot be ruled out that the real period is
twice shorter.  For $P=1.62$ day the likely reason of observed
variability would be binary nature of OGLEGC41. This star could
be either an eclipsing binary or an ellipsoidal variable. 
For $P=0.81$ day it could belong to BY~Dra or FK~Com type
variables. The relatively red color of OGLEGC41 
is consistent with the hypothesis that it belongs to spotted variables.  

The contact binary OGLEGC67 is located more than 0.5 mag to the red of
the base of the subgiant branch on the cluster CMD. 
It is most probably a foreground star.

The low amplitude variable OGLEGC68 is located about 2 mag above cluster
turnoff and 0.3 mag to the blue of the subgiant branch. We adopted for
it a period of 0.65 day but the light curve can be phased also with
a period of 1.3 day. 
For $P=0.65$ day the star could be a foreground spotted variable. 
For $P=1.3$ day it could be an ellipsoidal variable 
composed of subgiant and a star from an extended horizontal branch. 

\subsection{Cluster membership of the variables}
The $\omega $~Cen cluster is located at an intermediate galactic 
latitude of $b=+15$~deg. Therefore, one cannot assume  that 
all variables listed in Table 4 are cluster members. 
Figure 5 shows the period versus absolute magnitude diagram for 
10 SX~Phe stars listed in Table 5. The standard relations for the F-mode
pulsators (solid line) and the H-mode pulsators (dashed line) and for 
${\rm [Fe/H]=-2.2}$ (upper line) and ${\rm[Fe/H]=-0.7}$ (lower line) 
are also shown. The calibration $P-L-[{\rm Fe/H}]$ was taken from 
Nemec et al. (1994). We adopted for the cluster an apparent distance modulus  
$(m-M)_{\rm V}=13.86$  while calculating absolute 
magnitudes of SX~Phe stars. The assumed range of metallicities is based
on results published by Brown and Wallerstein (1993) and Vanture
et al. (1994).  The observed luminosities of reported SX~Phe 
stars are consistent  with the hypothesis that all of them are 
members of the cluster. 
 
We have applied the absolute brightness calibration established by 
Rucinski (1995) to calculate $M_{\rm V}$ for newly discovered 
contact binaries. Rucinski's calibration gives $M_{\rm V}$ as a function 
of period, unreddened color $(V-I)_{0}$ and metallicity:
\begin{eqnarray}
M_{\rm V}^{cal}=-4.43log(P)+3.63(V-I)_{0}-0.31-0.12\times [{\rm Fe/H}].
\end{eqnarray}
We adopted for all systems $[{\rm Fe/H}]=-1.6$ 
what is the mean metallicity for the cluster stars. 
Figure 6 shows the period versus an apparent distance modulus
diagram for contact binaries identified  in fields 5139D-F.
of $\omega $~Cen. Specifically the following systems listed in Table~6 
were considered to be "normal" W~UMa-type binaries: OGLEGC20, 44, 48,
49, 54, 56, 57, 58, 61, 64, 65 and 67. 
An apparent distance modulus was calculated for each 
system as a difference between its V magnitude and $M_{\rm V}^{cal}$. 
An apparent distance modulus for $\omega $~Cen is estimated at     
$(m-M)_{\rm V}=13.86$ (Nemec et al. 1994). The systems with mostly
deviating values of $(m-M)_{\rm V}$ are OGLEGC48, OGLEGC57
and OGLEGC67 (this system is not marked in Fig. 6). 
These binaries are probably 
a foreground variables. The remaining systems plotted in Fig. 6 
are likely members of the cluster.

Clearly some additional data are needed to establish with more confidence
which variables identified in our survey are members of $\omega $~Cen. 
We note that potential field interlopers should exhibit radial velocity
differing significantly from the velocity observed for the cluster. 
Radial velocity of $\omega $ Cen is +228 km/sec (eg. Zinn 1985 ) 
what is a much higher value than velocities observed for field stars 
from the cluster region. 
\section{The color-magnitude diagrams}
In Fig. 7 we show the $V$ vs. $V-I$  CMDs for fields 5139D-F. 
For each field the photometry is based on a single pair of 
"long" V\&I exposures. The CMD's based on "short" exposures 
are presented in Fig. 8. The frames used for construction of CMD's
shown if Figs. 7 and 8 are listed in Table 7.  
The presented CMDs were cleaned from stars of relatively poor photometry.

The presented CMDs show the same features which have been already 
noted in Paper I.  Several likely hot subdwarfs with $18.2<V<19.5$ 
and $(V-I)\approx 0.0$ can be noted. Also a numerous candidates 
for blue stragglers are visible, especially in fields 5139D and
5139E. The apparent main sequence of 
field stars can be traced from the base of the subgiant branch of
the cluster up to $V\approx 12.0$ what is the bright limiting magnitude
of our photometry. This sequence is probably related to the so called
"old disc" discussed recently by Jonch-Sorensen \& Knude (1994) and
Ng et al. (1995).   
All photometry presented in this section was submitted in tabular form 
to the editors of A\&A and is available in electronic form to all
interested readers (see Appendix A). The potential users of this photometry
should be aware about possibility of some systematic errors of 
the photometry. These errors are most likely to be significant for 
relatively faint stars. The CCD chip used for observations by the OGLE 
suffers from some nonlinearity. 
More details on this subject can be found in Paper I. 
\section{Blue stragglers}
The CMD's of all  monitored fields contain sizeable population of 
candidates for blue stragglers. Blue stragglers are known to be present 
in most well studied globular clusters (eg. Ferraro et al. 1995).
Several lines of evidence indicate that blue stragglers have higher 
masses than turnoff stars in their parent clusters. This supports
hypothesis, originally proposed by McCrea (1964) that
they are formed by merging of close binaries. In fact  a few dozen 
photometric binary blue stragglers. 
were discovered in globular clusters during  recent years 
(see Mateo (1996) for a recent summary). These are mostly W UMa-type
contact systems but some detached and semi-detached systems are
are also known. Another type of variable blue stragglers 
occurring in globulars are SX~Phe stars. The fraction of variables
seems to vary significantly from cluster to
cluster. For example about 30\% of blue stragglers in
NGC5053 are short-period variables (Nemec et al. 1995)
while no single variable could be identified among  27 blue stragglers
in NGC~6366 (Harris 1993). It is not clear for the moment how the
relative frequency of variable blue stragglers correlates with such
parameters as age and metallicity of a host cluster.

We used our data to estimate a relative frequency of contact binaries
and SX~Phe stars among $\omega $~Cen blue stragglers. It has to be noted
at this point that the definition of blue stragglers is somehow
flexible. It is unclear how to separate regions occupied by the turnoff
stars and faint blue stragglers. 
For fields 5139A, 5139BC, 5139C  and 5139D the following limits are adopted 
to define area occupied by blue stragglers: $16<V<18$, $0.30<V-I<0.70$.
For fields 5139E and 5139F the corresponding limits are: $16<V<18$,
$0.30<V-I<0.75$. The different definition of the red edge of the 
blue stragglers area reflects differences in the turnoff color 
observed for monitored fields. These differences may be possibly caused by
the patchy reddening across the cluster face. 
The data presented in Fig. 7 and the data presented in Fig. 12 from
Paper I were used to determine a number of blue stragglers in each of
monitored fields. Note that presented CMD's are highly incomplete as we
plotted only stars with relatively reliable photometry. The degree of
incompleteness varies with changing  crowding. 
Table 8 gives a number of selected blue stragglers candidates. 
Only objects whose light curves were analyzed for variability 
were retained in this statistic. The number of variable blue 
stragglers which were included in presented CMD's is also given 
in Table 8. All counted objects are blue straggles according
to adopted criteria on color and magnitude\footnote{In fact 
we used for variables colors listed in Tables 5 and 6 (Tables 6 and 7 
for fields studied in Paper I). The position marked on presented CMD's 
are based on a single pair V \& I of frames. Therefore 
they may be systematically in error in case of variable stars.}.
Note that some detected variable blue stragglers were dropped from 
the statistic. These were stars with formally poor photometry 
in the data sets used for plotting the CMD's.
The fields 5139A and 5139C-F overlap with field 5139BC. Moreover, field 
5139F overlaps with fields 5130E and 5139C. 
The statistics given in Table 8 applies to the non-overlapping
parts of surveyed fields. The effective area of all fields are given in
the second column of Table 8. 
We counted a total of 200 candidate blue stragglers. This number
includes 24 SX~Phe stars and 11 binaries (9 EW and 2 EB/EA systems). 
We showed that most of detected variables are very likely  cluster
members. 
The lower limits on the relative frequency of SX~Phe stars and eclipsing
binaries among cluster blue straggles are $(24/200)=0.12$ and
$11/200=0.055$, respectively.
It is difficult to asses precisely what fraction of candidate 
blue stragglers are field interlopers. However, we note that 
the surface density of possible blue stragglers is by a factor of
$(92/10)\times 0.75=6.0$  higher for the central field 5139BC than for the
outermost field 5139F. It may be concluded that most of selected blue
stragglers are indeed members of the cluster. 
Our data indicate that $\omega $~Cen contains the largest number of blue
stragglers among all well studied galactic globular clusters (eg.
Ferraro et al. 1995). 

\section{Summary}
This section summarizes results presented above as well as those
described in Paper~I.
We surveyed a total of six fields covering most of the central part 
of $\omega $~Cen and identified 70 periodic variables.
This sample includes 5 previously known variables. RR~Lyr stars are excluded 
from these statistics -- a separate paper devoted to them is in
preparation.  
The most interesting result is identification of 3 detached eclipsing 
systems which are located at the base of the subgiant branch of the 
cluster.  These are OGLEGC15, OGLEGC17 and OGLEGC51.
Further observations of these systems can lead to direct determination of
masses and radii of turnoff stars in $\omega $~Cen. An accurate
determination of the cluster distance would be also obtained from 
such data. Another clearly detached binaries discovered in our survey
are OGLEGC14 and OGLEGC52. The former one is a  blue straggler (assuming
it is a cluster member) while the latter is located slightly above 
the horizontal branch of the cluster. Determination of parameters for
components of OGLEGC52 would provide information about masses of
horizontal branch stars in $\omega$~Cen. 

We obtained light curves for 34 SX~Phe stars. All of them are 
located among blue stragglers and are likely members of the 
cluster. 20 out of 34 of identified SX~Phe stars shows 
variability with a full range not exceeding 0.1 mag in the V-band.
Detected SX~Phe stars constitute about 11\% of the blue straggler 
population of $\omega $~Cen. 

Several contact or nearly-contact binaries were identified. Most of
them are located among blue stragglers but some systems are
present among upper main-sequence and turnoff stars. 
Most of identified contact binaries are likely members of the cluster.

Three low-amplitude variables with periods from the range 19-33 days
were identified on the subgiant branch (OGLEGC22, OGLEGC35 and
OGLEGC69). These stars belong most likely to the so called "spotted" 
variables. Another likely variable of this type is OGLEGC30 
whose position on the CMD is unknown due to lack of information 
about its color.

\section{Appendix A}
Tables containing light curves of all variables discussed in this
paper as well as tables with VI photometry presented in Figs. 7 and 8
are published
by A\&A at the centre de Donn\'{e}es de Strasbourg, where they are available
in electronic form: See the Editorial in A\&A 1993, Vol. 280, page E1.
Equatorial coordinates are given for all stars included in these tables. 
We have also submitted the V-filter "template" images of
fields 5139D-F. These images can be used for identification of all variables 
discussed in this paper as well as for identification of all stars 
for which we provide VI photometry.  
\acknowledgements
This project was supported by NSF grants AST 92-16494 and AST-9313620
to Bohdan Paczynski and AST 92-16830 to George Preston. MK, MS and AU
were supported by the Polish KBN grant 2P03D02908.
JK was supported by the Polish KBN grants 2P03D00808 and BST 501/17/95.

\clearpage
\setcounter{table}{2}
% Table3
\begin{table}
\begin{center}
\caption[]{
Basic statistical data about stars from fields 5139D-F examined 
for variability. The data are given for bins 0.5 mag wide.
Columns 2, 4 and 6 give {\it median} value of {\it rms} for
a given bin. Columns 3, 5  and 7 give numbers of stars examined 
for variability.
}
%\begin{tabular}{lllllll}
\begin{tabular}{ccrrcrr}
\tableline
\tableline
    &D    &     &   E    &    & F &     \\
V   &$<rms>$  &  N  & $<rms>$ & N  &rms & N \\
\tableline
\tableline
14.5  &   0.015  & 125   & 0.019  & 175  &  0.014& 66 \\
15.0  &   0.010  & 187   & 0.012  & 231  &  0.011& 107 \\
15.5  &   0.010  & 144   & 0.013  & 219  &  0.011& 92 \\
16.0  &   0.011  & 215   & 0.014  & 270  &  0.012& 129 \\
16.5  &   0.013  & 305   & 0.018  & 418  &  0.015& 152 \\
17.0  &   0.017  & 392   & 0.021  & 579  &  0.019& 238 \\
17.5  &   0.020  & 923   & 0.027  & 1340 &  0.023& 547 \\
18.0  &   0.026  & 2129  & 0.036  & 3040 &  0.029& 1115 \\
18.5  &   0.036  & 3451  & 0.047  & 5185 &  0.041& 1913 \\
19.0  &   0.043  & 1540  & 0.059  & 2200 &  0.056& 2550 \\
\tableline
\end{tabular}
\end{center}
\end{table}
\clearpage
%Table 4
\setcounter{table}{3}
\begin{table}
%\scriptsize
\caption[]{Rectangular and equatorial coordinates for variables
identified in $\omega$~Cen. The X and Y coordinates give
positions of variables on the template images (see text for details). 
The last column gives
alternative names for two variables which were previously known. 
Variability of OGLEGC17 and OGLEGC20 was already reported in Paper I.
 }
\begin{flushleft}
\begin{tabular}{lrrrrrr}
\hline\noalign{\smallskip}
%\hline\noalign
Name & X & Y & RA(2000) & Dec(2000) & Field & Other\\
     &   &   &  (h:m:s) & (deg:':'')        &       & Name \\
\hline\noalign{\smallskip}
OGLEGC17 &  113.69& 1800.14 & 13:27:21.87& -47:23:32.2 &5139E & \\
OGLEGC20 &   99.82& 1573.85 & 13:27:21.71& -47:37:19.6 &5139F & \\
OGLEGC41 &  683.38&  803.52 & 13:25:08.91& -47:30:24.9 &5139D & \\
OGLEGC42 &  551.31& 1493.66 & 13:25:01.10& -47:25:29.9 &5139D & E39\\
OGLEGC43 & 1002.56& 1873.04 & 13:25:19.23& -47:22:31.3 &5139D & \\
OGLEGC44 & 1751.23& 1593.30 & 13:25:52.34& -47:24:07.9 &5139D & \\
OGLEGC45 & 1770.24& 1620.64 & 13:25:53.08& -47:23:55.4 &5139D & \\
OGLEGC46 &  341.21&  975.77 & 13:27:34.05& -47:29:23.4 &5139E & \\
OGLEGC47 &  215.52&  987.21 & 13:27:28.68& -47:29:22.3 &5139E & \\
OGLEGC48 &  248.03& 1405.29 & 13:27:28.78& -47:26:19.7 &5139E & \\
OGLEGC49 &  448.23& 1741.81 & 13:27:36.24& -47:23:47.5 &5139E & \\
OGLEGC50 &  780.64&  596.16 & 13:27:53.92& -47:31:54.4 &5139E &NJL220 \\
OGLEGC51 &  596.32&  901.86 & 13:27:45.13& -47:29:47.6 &5139E & \\
OGLEGC52 &  607.51& 1403.16 & 13:27:44.06& -47:26:09.7 &5139E & \\
OGLEGC53 & 1497.66&  999.54 & 13:28:23.28& -47:28:37.0 &5139E & \\
OGLEGC54 & 1761.09&  167.23 & 13:28:37.19& -47:34:29.0 &5139E & \\
OGLEGC55 & 1742.23& 1286.89 & 13:28:32.82& -47:26:24.7 &5139E & \\
OGLEGC56 & 1052.01&  357.10 & 13:28:06.77& -47:45:33.8 &5139F & \\
OGLEGC57 & 1141.94& 1016.56 & 13:28:08.31& -47:40:45.2 &5139F & \\
OGLEGC58 &  584.86& 1662.45 & 13:25:02.01& -47:24:15.7 &5139D & \\
OGLEGC59 & 1589.98& 1281.47 & 13:25:46.37& -47:26:28.3 &5139D & \\
OGLEGC60 &  227.27&  110.09 & 13:27:31.80& -47:35:42.6 &5139E & \\
OGLEGC61 &  228.86& 1350.74 & 13:27:28.14& -47:26:44.0 &5139E & \\
OGLEGC62 &  246.23& 1035.29 & 13:27:29.84& -47:29:00.5 &5139E & \\
OGLEGC63 &   81.06& 1204.51 & 13:27:22.31& -47:27:52.0 &5139E & \\
OGLEGC64 &  233.51& 1222.93 & 13:27:28.73& -47:27:39.4 &5139E & \\
OGLEGC65 & 1825.39& 1621.01 & 13:28:35.29& -47:23:57.3 &5139E & \\
OGLEGC66 &   51.12& 1774.18 & 13:27:18.97& -47:35:54.4 &5139F & \\
OGLEGC67 &  610.93& 1091.16 & 13:25:04.92& -47:28:22.5 &5139D & \\
OGLEGC68 & 1410.64& 1412.39 & 13:25:38.23& -47:25:37.5 &5139D & \\
OGLEGC69 & 1956.06&  832.21 & 13:26:03.58& -47:29:30.6 &5139D & \\
OGLEGC70 &  355.10& 1525.39 & 13:24:52.62& -47:25:22.2 &5139D & \\
\hline\noalign{\smallskip}
\end{tabular}
\end{flushleft}
\end{table}
\clearpage
%Table 5
\setcounter{table}{4}
\begin{table}
\caption[]{
Light-curve parameters for SX~Phe stars from fields D-F in $\omega $~Cen.
$A_{V}$ is the range of observed variations in the $V$ band.
$(V-I)$ is the observed color at the maximum light. 
The period is given in days.
}
\begin{flushleft}
\begin{tabular}{rrrrrr}
\hline\noalign{\smallskip}
Name & Period& $<V>$ & $V{\rm max}$ & $A_{V}$ & $(V-I)$ \\
\hline\noalign{\smallskip}
OGLEGC42 &  0.05740 & 17.00 & 16.86&	0.23 & 0.51\\
OGLEGC45 &  0.06560 & 16.84 & 16.69&	0.24 & 0.57\\
OGLEGC46 &  0.04080 & 17.40 & 17.36&	0.08 & 0.58\\
OGLEGC50 &  0.04718 & 17.04 & 16.75&	0.48 & 0.49\\
OGLEGC59 &  0.03495 & 17.51 & 17.46&	0.10 & 0.53\\
OGLEGC60 &  0.04063 & 17.47 & 17.43&	0.06 & 0.63\\
OGLEGC62 &  0.04662 & 17.46 & 17.43&	0.06 & 0.67\\
OGLEGC63 &  0.03997 & 17.32 & 17.28&	0.08 & 0.50\\
OGLEGC66 &  0.03749 & 17.53 & 17.49&	0.07 & 0.52\\
OGLEGC70 &  0.04626 & 17.08 & 17.05&	0.05 & 0.56\\
\hline\noalign{\smallskip}
\end{tabular}
\end{flushleft}
\end{table}
%\clearpage
%Table 6
\setcounter{table}{5}
\begin{table}
\begin{center}
\caption[]{
Light-curve parameters for eclipsing binaries discovered in fields
D-F in $\omega $~Cen. 
$(V-I)$ is the observed color at the maximum light.
$T_{0}$ is the time of minimum light.
}
\begin{tabular}{rllrrrr}
\tableline
\tableline
Name & Type  & Period&$V_{\rm max}$ & $V_{\rm min}$ & $V-I$ &$T_{0}$ HJD \\
     &       & (days)&              &               &     & 2449000+   \\
\tableline
\tableline
OGLEGC17& EA  &  2.46655& 17.29& 17.67& 0.99& 818.7271\\
OGLEGC20& EW  &  0.34183& 16.60& 16.75& 0.41& 818.8224\\
OGLEGC41& Ell/BY & 1.62/0.81 & 16.62& 16.71& 1.48& 516.3300\\
OGLEGC43& EW  &  1.09644& 18.57& 18.84& 0.99& 516.4300\\
OGLEGC44& EW  &  0.29631& 17.26& 17.49& 0.60& 515.7007\\
OGLEGC47& EA  &  1.18870& 16.70& 17.58& 0.67& 819.2760\\
OGLEGC48& EW  &  0.33188& 16.37& 16.48& 0.61& 818.7239\\
OGLEGC49& EB  &  0.36627& 18.57& 19.30& 0.92& 818.5690\\
OGLEGC51& EA  &  4.64042& 17.08& 17.34& 0.98& 820.7450\\
OGLEGC52& EA  &  1.16807& 14.12& 14.83& 0.49& 818.8469\\
OGLEGC53& EB  &  0.49519& 18.38& 19.00& 0.92& 818.8560\\
OGLEGC54& EW  &  0.28377& 18.94& 19.18& 1.09& 818.7926\\
OGLEGC55& EB  &  0.40389& 17.21& 17.85& 1.05& 818.7440\\
OGLEGC56& EW  &  0.28118& 19.23& 19.73& 1.12& 818.7473\\
OGLEGC57& EW  &  0.41783& 16.95& 17.09& 1.03& 818.7016\\
OGLEGC58& EW  &  0.41577& 16.82& 16.92& 0.64& 515.6599\\
OGLEGC61& EW  &  0.63093& 16.19& 16.23& 0.52& 818.9218\\
OGLEGC64& EW  &  0.38505& 16.47& 16.56& 0.48& 818.8974\\
OGLEGC65& EW  &  0.51224& 17.16& 17.23& 0.61& 818.8738\\
OGLEGC67& EW  &  0.25064& 17.52& 17.69& 1.70& 515.7382\\
OGLEGC68& Ell/RS? &  0.65469& 15.78& 15.82& 0.71& 818.4927\\
OGLEGC69& RS? & 19.18   & 15.71& 15.77& 1.13& 820.04\\
\tableline
\end{tabular}
\end{center}
\end{table}
\clearpage
%Table 7
\setcounter{table}{6}
\begin{table}
\begin{center}
\caption[]{
List of frames used for construction of CMDs shown in Figs. 7 and 8.
}
\begin{tabular}{rrrrr}
\tableline
\tableline
Frame & Field &$T_{\rm exp}$ & Filter & FWHM \\
      &       & sec          &        & arcsec \\
\tableline
mr10525 & 5139D& 420  &V & 1.01 \\
mr10532 & 5139D& 300  &I & 0.97 \\
mr10184 & 5139D&  90  &V & 0.94 \\
mr10188 & 5139D&  60  &I & 1.14\\
mr10524 & 5139E& 420  &V & 1.01 \\
mr10531 & 5139E& 300  &I & 0.99 \\
mr10185 & 5139E&  90  &V & 0.98 \\
mr10187 & 5139E&  60  &I & 0.99\\
mr11042 & 5139F& 420  &V & 1.00 \\
mr11447 & 5139F& 300  &I & 1.01 \\
mr10979 & 5139F&  90  &V & 0.94 \\
mr10998 & 5139F&  76  &I & 1.07\\
\tableline
\end{tabular}
\end{center}
\end{table}
\clearpage
%Table 8
\setcounter{table}{7}
\begin{table}
\begin{center}
\caption[]{
Number of blue straggler candidates and number of
variable blue stragglers in 6 $\omega $~Cen fields.
The second column gives relative area of a given field in units
of area of the field 5139BC. 
N(SX) and N(ecl) denote numbers of SX~Phe stars and eclipsing 
binaries, respectively. The last row summarizes data for all fields}
\begin{tabular}{lrrrr}
\tableline
\tableline
Field &Area & N(BS) & N(SX) & N(ecl) \\
\tableline
%5139A 27 7 2
%5139BC 92 6 5 
%5139C 28 5 2
%5139D 28 4 1 	
%5139E 36 3 3
%5139F 16 2  1
%5139A-F 
5139A   &0.94 &25  &7 & 1\\
5139BC  &1.0  &92  &6 & 5 \\
5139C   &0.87 &20  &4 & 1\\
5139D   &0.92 &26  &4 & 1 \\	
5139E   &0.93 &27  &3 & 3\\
5139F   &0.75 &10  &0 & 0\\
5139A-F &5.41 &200 &24& 11\\
\tableline
\end{tabular}
\end{center}
\end{table}
\clearpage
%
%fig1
\setcounter{figure}{0}
\begin{figure}
\epsscale{1.0}
\plotone{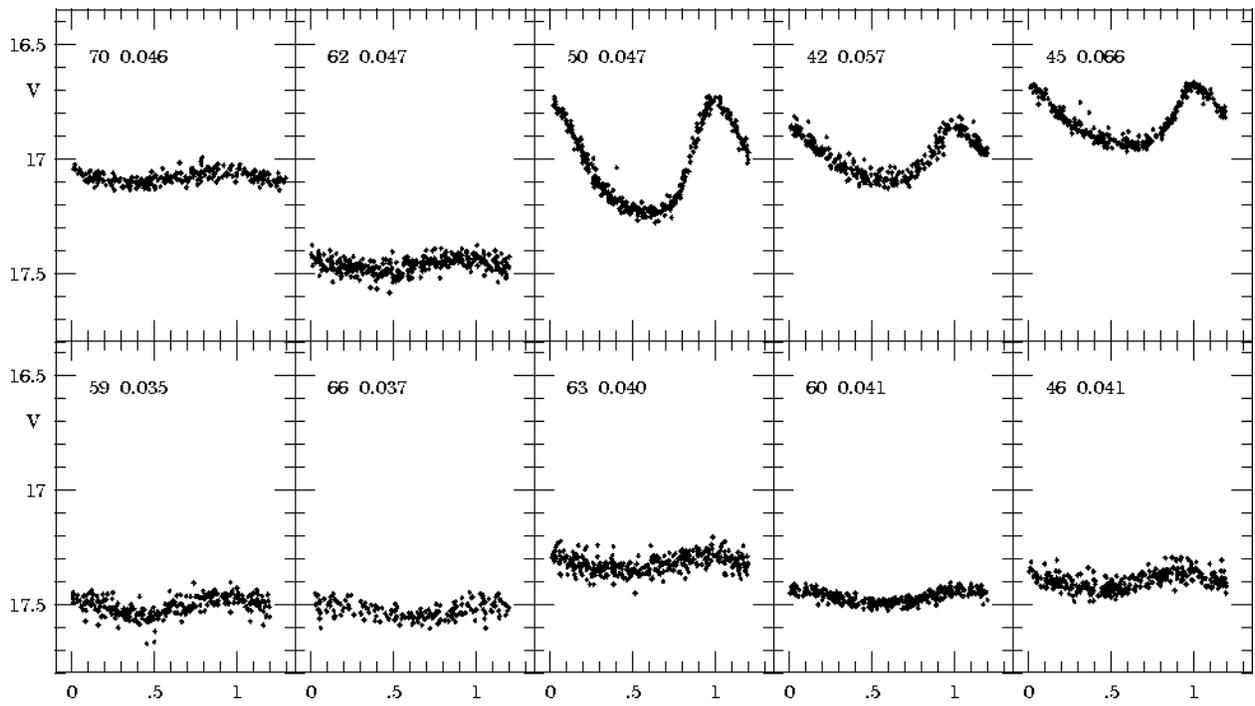}
\caption[]{Phased $V$ light curves for 10 SX~Phe stars identified 
in fields 5139D-F. Note the same height of all boxes.  Variables 
are sorted according to their periods: inserted labels give names of
variables and their periods in days.
}
\end{figure}
\clearpage
%fig2
\setcounter{figure}{1}
\begin{figure}
\epsscale{1.0}
\plotone{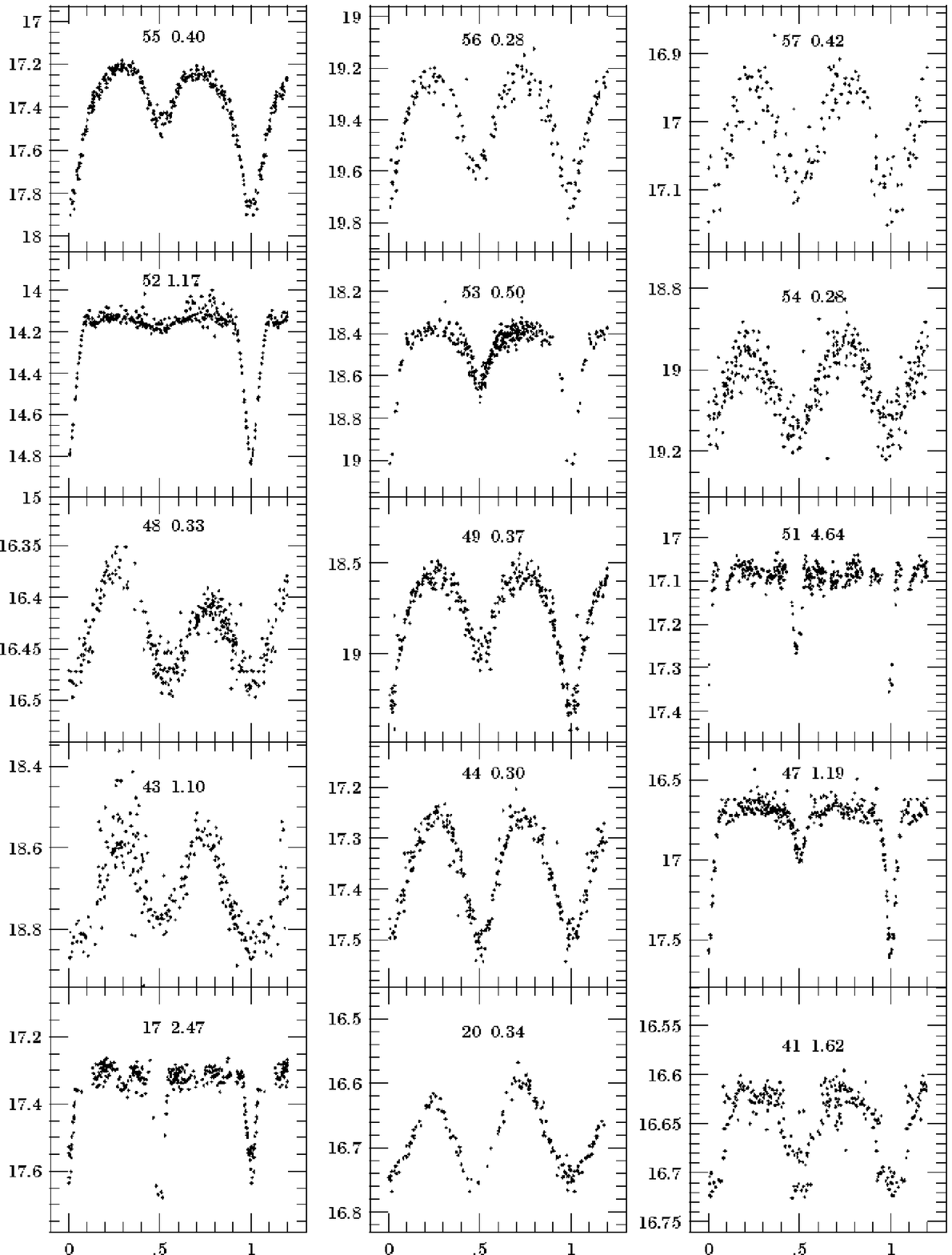}
\caption[]{Phased $V$ light curves for non-pulsating  variables
identified in fields 5139D-F.
Inserted labels give names of variables and
their periods in days.
}
\end{figure}
\clearpage
%
%fig3
\setcounter{figure}{2}
\begin{figure}
\epsscale{1.0}
\plotone{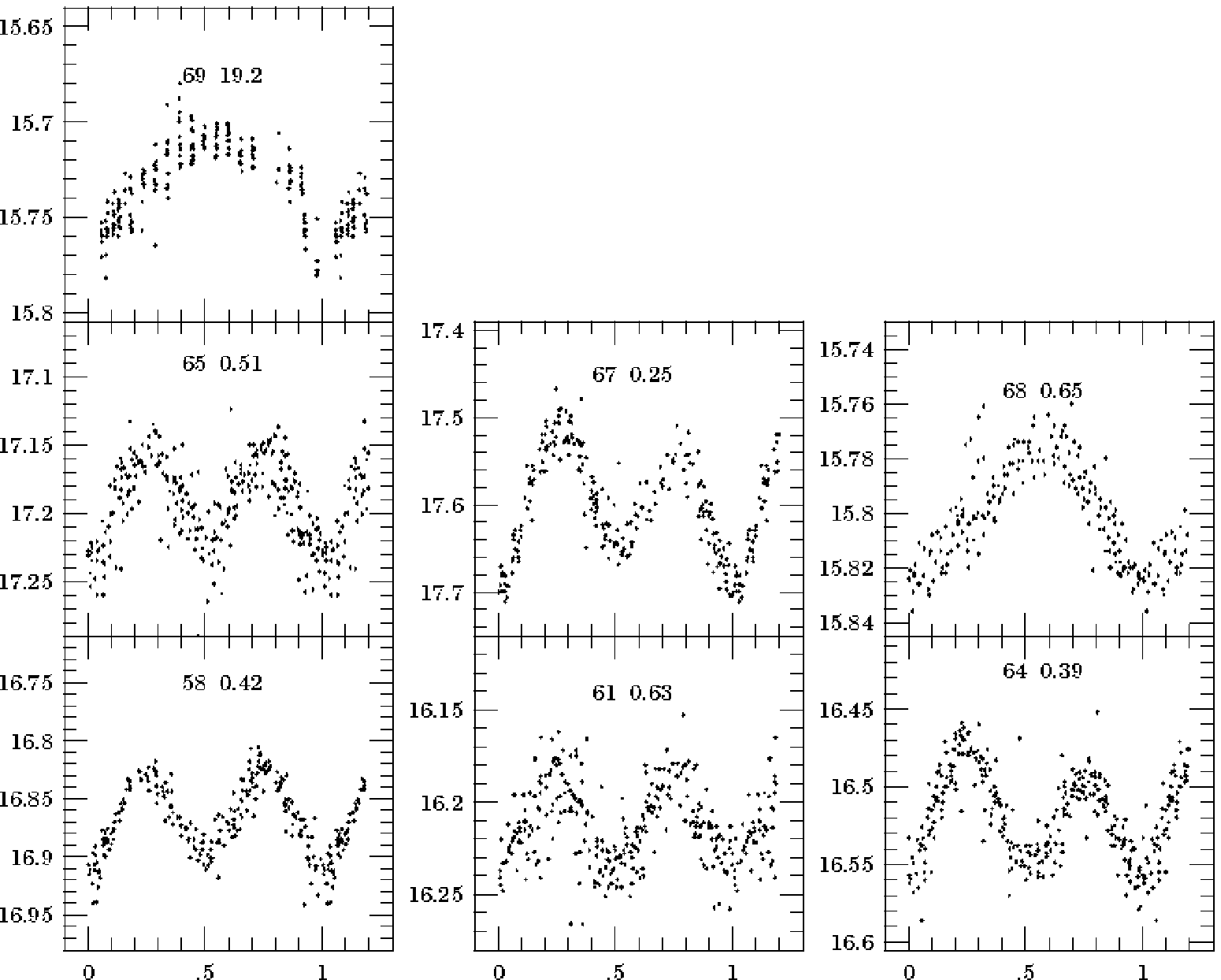}
\caption[]{
Phased $V$ light curves for non-pulsating variables
identified in fields 5139D-F. Inserted labels give names of variables and
their periods in days.
}
\end{figure}
\clearpage
%
%fig4
\setcounter{figure}{3}
\begin{figure}
\epsscale{1.0}
\plotone{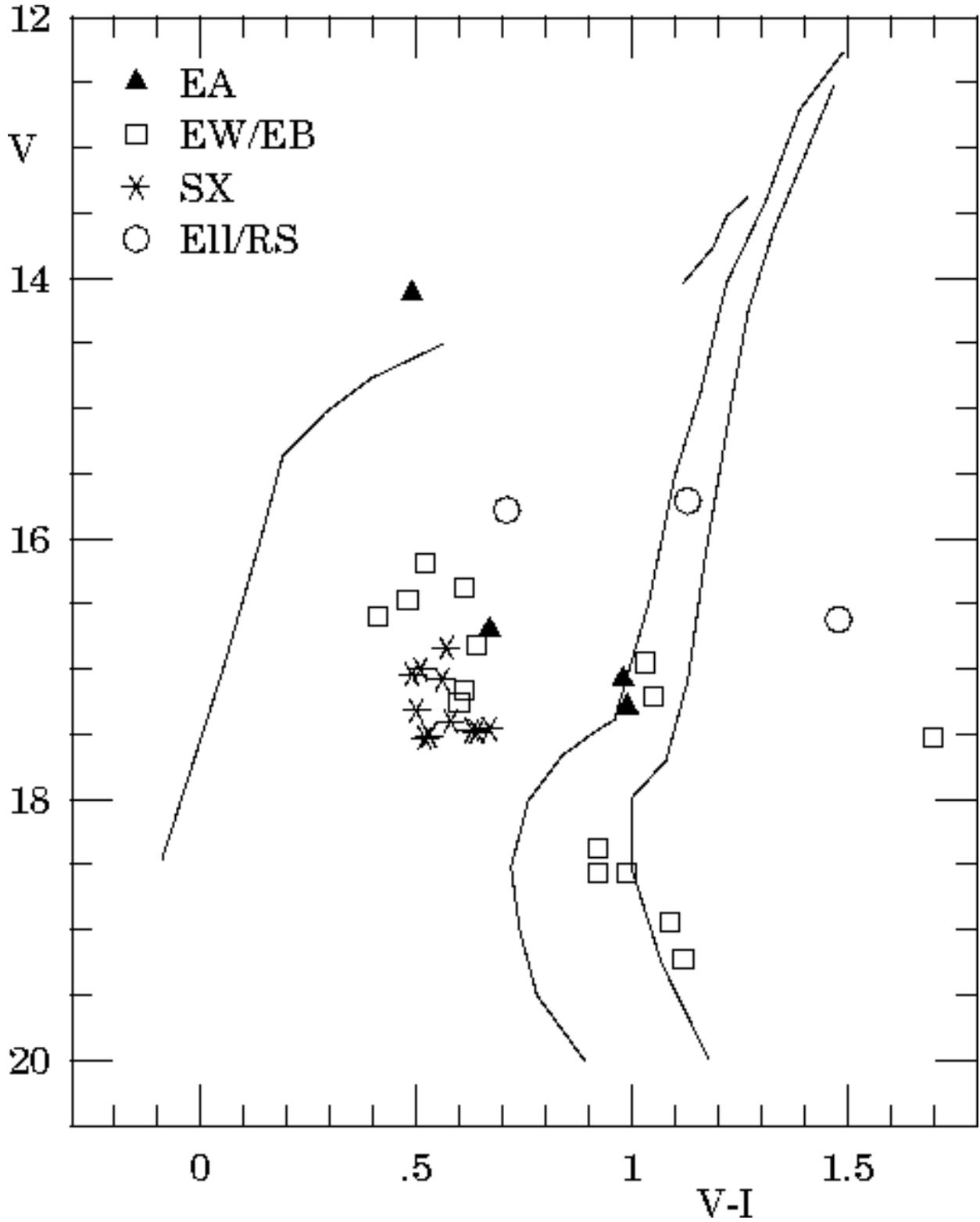}
\caption[]{The schematic CMD for $\omega $~Cen with positions
of variables from fields D-F marked. The triangles represent EA-type 
eclipsing binaries, the squares EW/EB-type binaries, the asterisks 
SX~Phe stars and the open circles probable ellipsoidal or 
spotted variables. 
}
\end{figure}
\clearpage
%fig5
\setcounter{figure}{4}
\begin{figure}
\epsscale{1.0}
\plotone{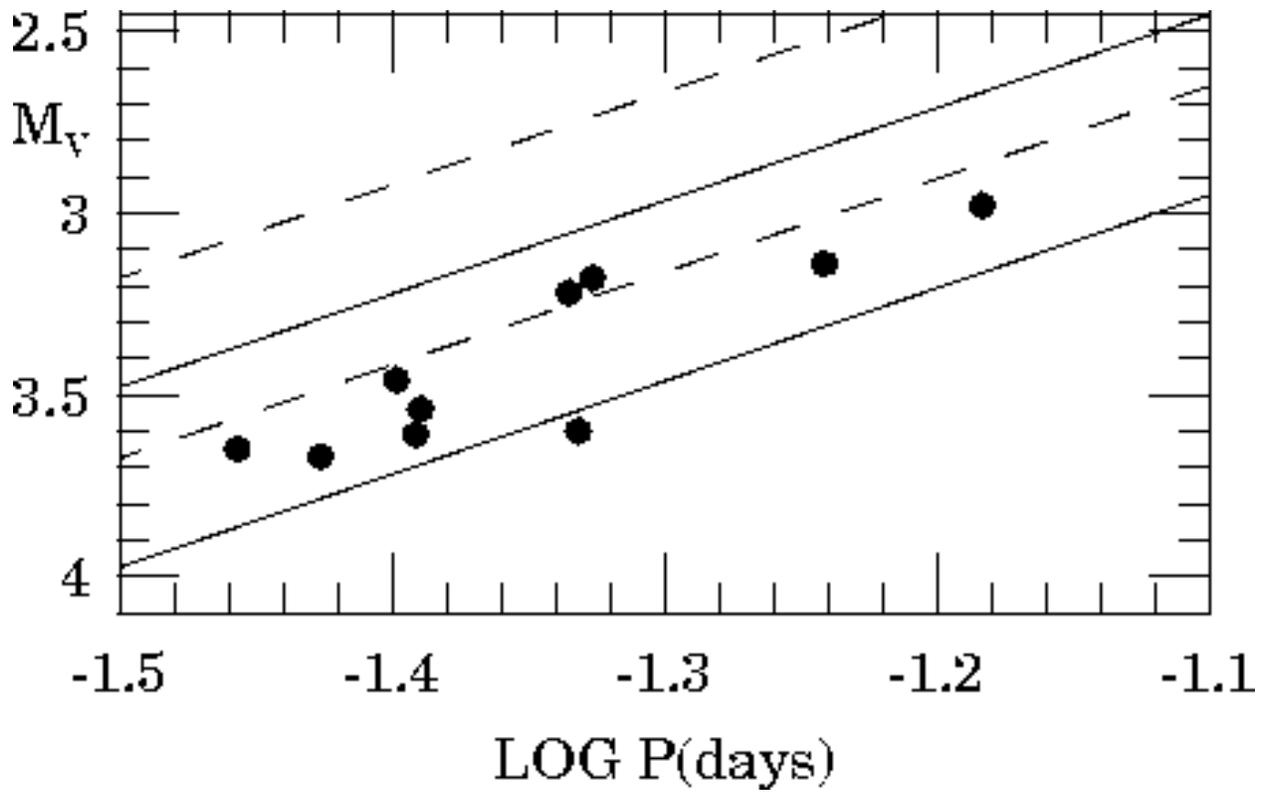}
\caption[]{
Period vs. absolute magnitude diagram for SX~Phe stars from the 
field of $\omega$~Cen.
}
\end{figure}
\clearpage
%fig6
\setcounter{figure}{5}
\begin{figure}
\epsscale{1.0}
\plotone{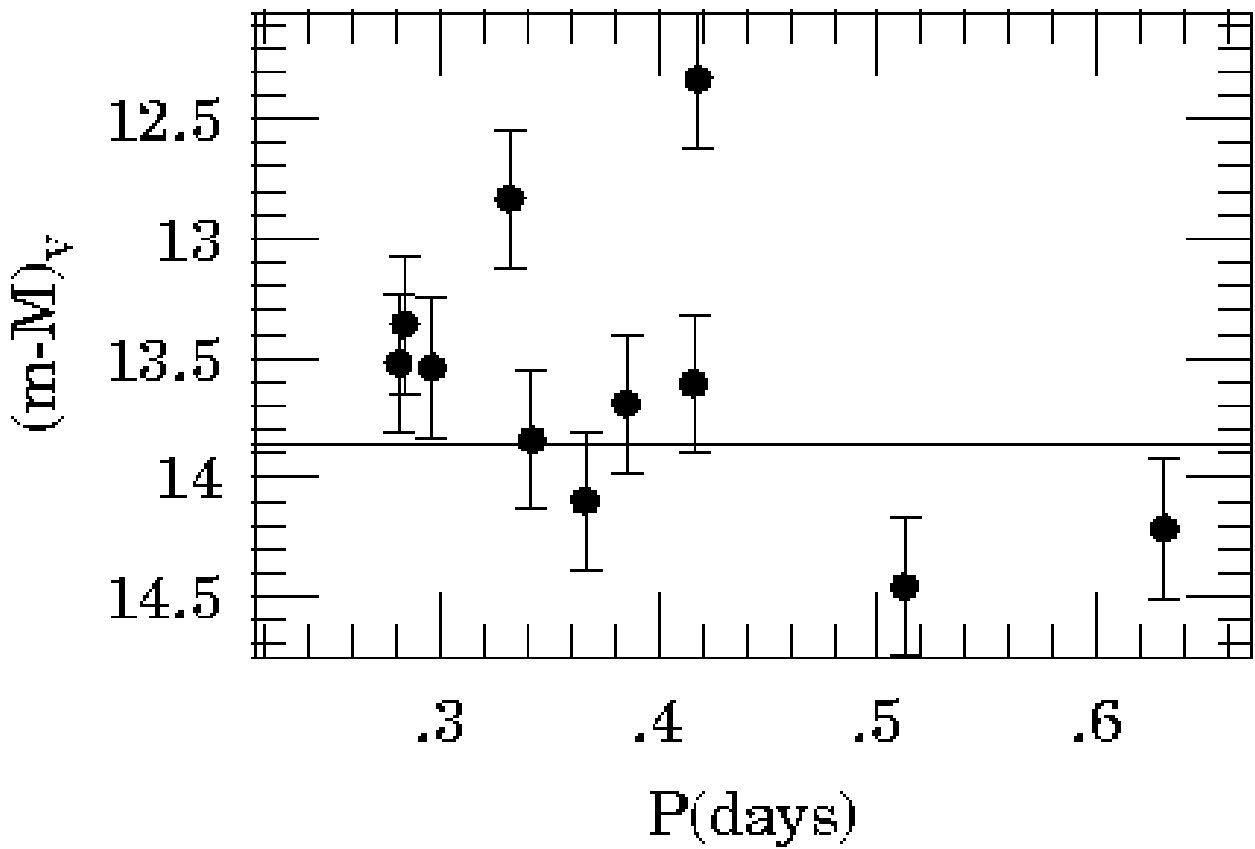}
\caption[]{Period, apparent distance modulus diagram for contact 
binaries from the field of $\omega$~Cen. A horizontal line at 
$(m-M)_{\rm V}=13.86$ corresponds to distance modulus of the cluster.
Error bars correspond to the formal uncertainty of absolute magnitudes
derived using Rucinski's (1995) calibration.
}
\end{figure}
\clearpage
%
%fig7
\setcounter{figure}{6}
\begin{figure}
\epsscale{0.7}
\plotone{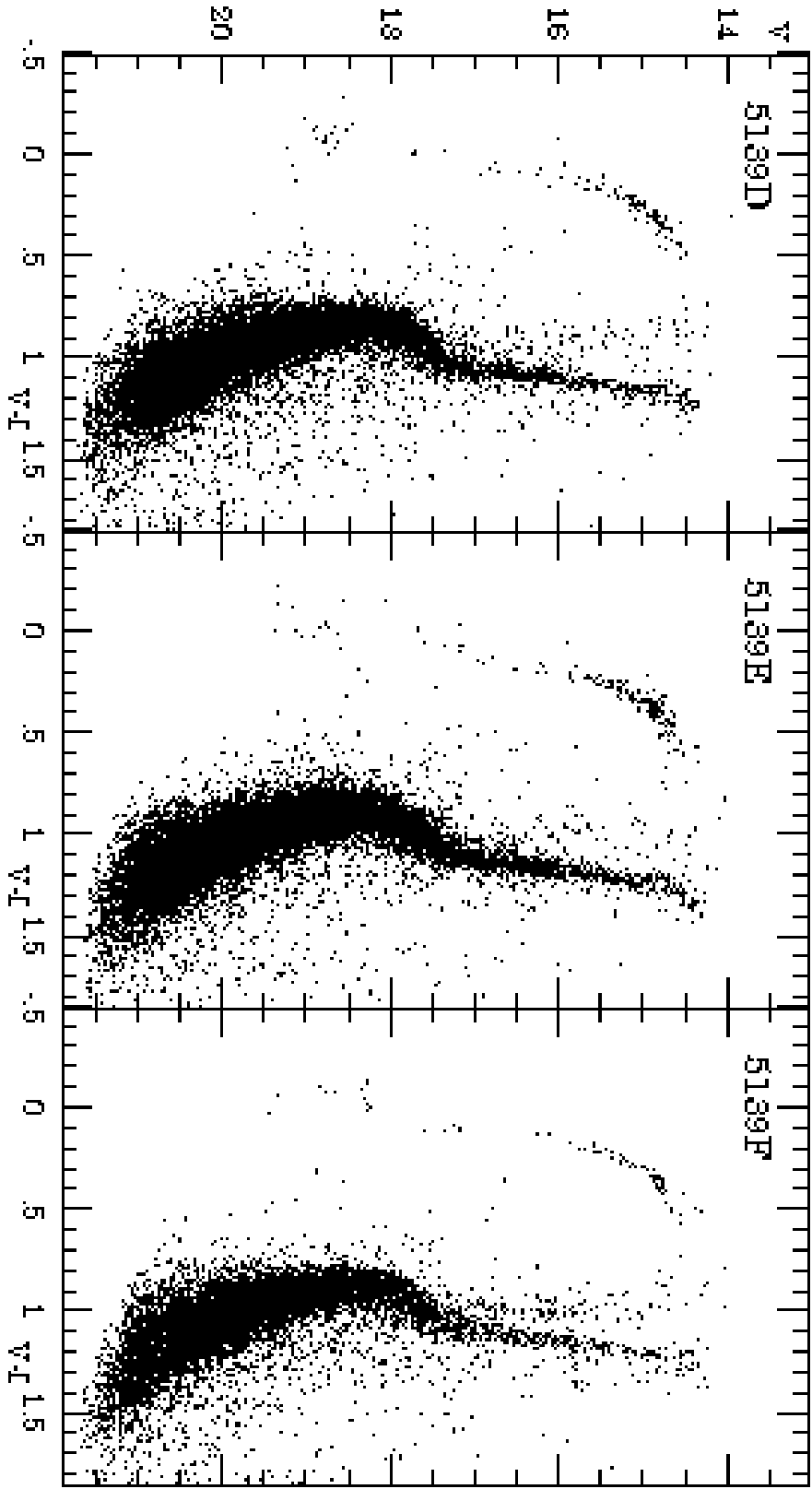}
\caption[]{
The CMDs for fields 5139D (left), 5139E (center) and 5139F (right).
}
\end{figure}
\clearpage
%fig8
\setcounter{figure}{7}
\begin{figure}
\epsscale{0.7}
\plotone{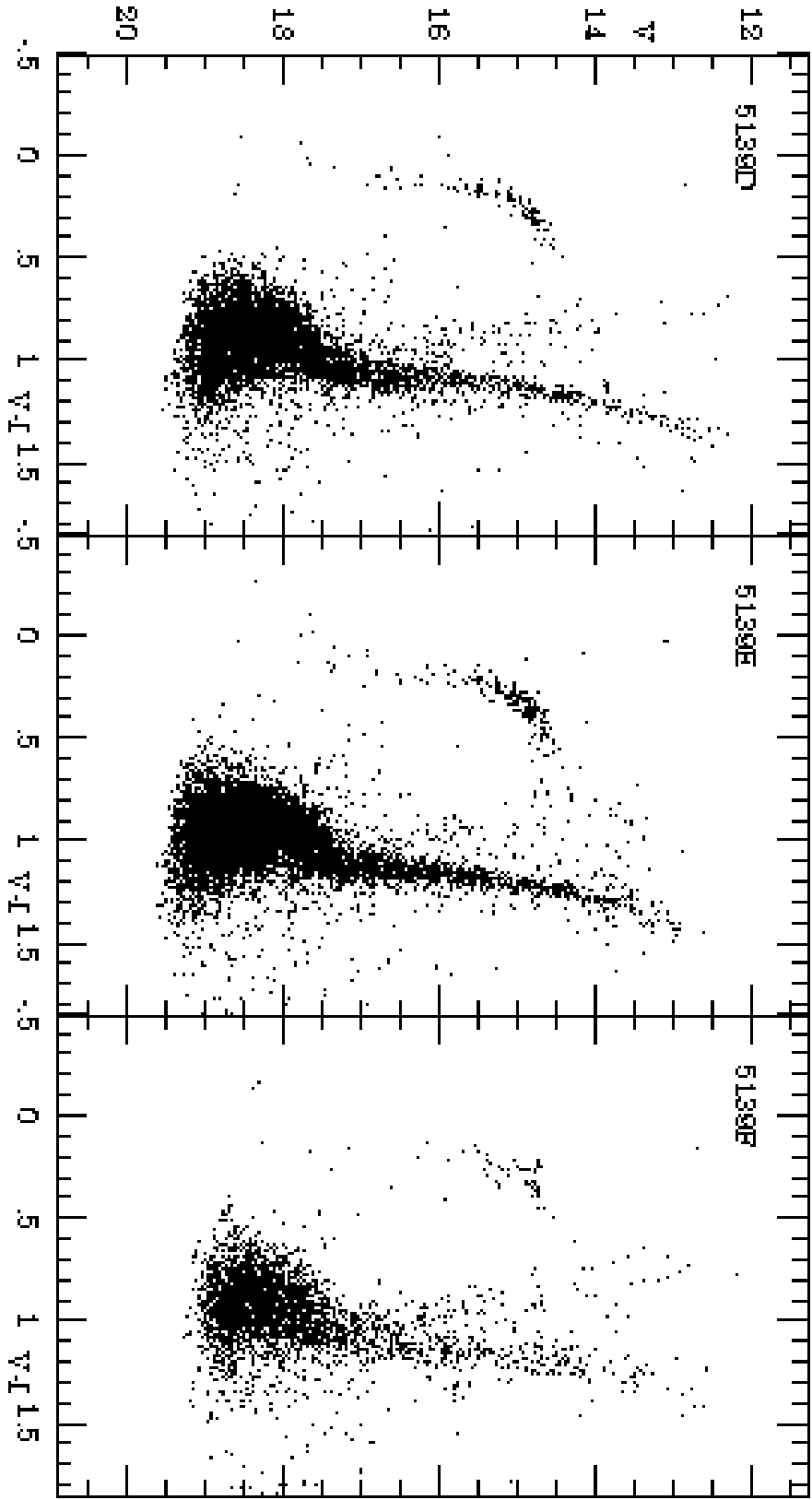}
\caption[]{
The CMDs for fields 5139D (left), 5139E (center) and 5139F (right)
based on shortly exposed frames. 
}
\end{figure}
\end{document}